\documentclass{article}

\usepackage{arxiv}

\usepackage[utf8]{inputenc} 
\usepackage[T1]{fontenc}    
\usepackage{hyperref}       
\usepackage{url}            
\usepackage{booktabs}       
\usepackage{amsfonts}       
\usepackage{nicefrac}       
\usepackage{microtype}      
\usepackage{lipsum}		
\usepackage{graphicx}
\usepackage{doi}
\usepackage{authblk} 
\usepackage{enumitem}
\usepackage{xurl}
\usepackage{multirow}
\usepackage[numbers]{natbib}
\usepackage{xcolor}

\title{VIEWER: an extensible visual analytics framework for enhancing mental healthcare}

\author[1]{Tao Wang, PhD}
\author[2]{David Codling, MB, MSc}
\author[1]{Yamiko Joseph Msosa, PhD}
\author[2]{Matthew Broadbent, MA}
\author[2]{Daisy Kornblum, PhD}
\author[1,2]{Catherine Polling, MB, PhD}
\author[1]{Thomas Searle, PhD}
\author[2]{Claire Delaney-Pope, MBA}
\author[2]{Barbara Arroyo, MB, MS}
\author[2]{Stuart MacLellan, MBA}
\author[2]{Zoe Keddie, PhD}
\author[2]{Mary Docherty, MB, MA}
\author[1]{Angus Roberts, PhD}
\author[1,2]{Robert Stewart, MD}
\author[3]{Philip McGuire, MD}
\author[1,4,5]{Richard Dobson, PhD}
\author[2]{Robert Harland, MB}

\affil[1]{Institute of Psychiatry, Psychology and Neuroscience, King's College London, UK}
\affil[2]{South London and Maudsley NHS Foundation Trust, UK}
\affil[3]{Department of Psychiatry, University of Oxford, UK}
\affil[4]{Institute of Health Informatics, University College London, UK}
\affil[5]{Health Data Research UK, UK}

\renewcommand{\shorttitle}{\textit{VIEWER}: an extensible visual analytics framework for enhancing mental healthcare}

\hypersetup{
pdftitle={A template for the arxiv style},
pdfsubject={q-bio.NC, q-bio.QM},
pdfauthor={David S.~Hippocampus, Elias D.~Striatum},
pdfkeywords={First keyword, Second keyword, More},
}

\begin{document}
\maketitle

\begin{abstract}
\textbf{Objective}: {A proof-of-concept study aimed at designing and implementing VIEWER, a versatile toolkit for visual analytics of clinical data, and systematically evaluating its effectiveness across various clinical applications while gathering feedback for iterative improvements.} 

\textbf{Materials and Methods:} {VIEWER is an open-source and extensible toolkit that employs natural language processing and interactive visualisation techniques to facilitate the rapid design, development, and deployment of clinical information retrieval, analysis, and visualisation at the point of care. Through an iterative and collaborative participatory design approach, VIEWER was designed and implemented in \textcolor{black}{one of the UK's largest NHS mental health Trusts}, where its clinical utility and effectiveness were assessed using both quantitative and qualitative methods.} 

\textbf{Results:} {VIEWER provides interactive, problem-focused, and comprehensive views of longitudinal patient data \textcolor{black}{(n=409,870)} from a combination of structured clinical data and unstructured clinical notes. Despite a relatively short adoption period and users' initial unfamiliarity, VIEWER significantly improved performance and task completion speed compared to the standard clinical information system. \textcolor{black}{More than 1,000 users and partners in the hospital tested and used VIEWER}, reporting high satisfaction and expressed strong interest in incorporating VIEWER into their daily practice.} 
        
\textbf{Discussion:} {VIEWER provides a cost-effective enhancement to the functionalities of standard clinical information systems, with evaluation offering valuable feedback for future improvements.} 
       
\textbf{Conclusion:} {VIEWER was developed to improve data accessibility and representation across various aspects of healthcare delivery, including population health management and patient monitoring. The deployment of VIEWER highlights the benefits of collaborative refinement in optimizing health informatics solutions for enhanced patient care.} 

\end{abstract}


\section{Introduction}

As the volume of information in patient records continues to grow, clinicians face an overwhelming amount of complex raw data, which can exceed the capacity of human cognition to process without error \cite{mcdonald1976protocol}. 
This issue of information overload has been widely noted across various care settings, including both primary and secondary care, which may result in diagnostic or treatment errors \cite{singh2013information,nijor2022patient}.
Moreover, the growing demand for health information exchange among healthcare providers, and between providers and partners such as public health agencies and regulators, has made patient data review and administrative reporting increasingly complex and time-consuming \cite{adler2011survey}. 
This is particularly evident in mental healthcare, where patients often require ongoing mental health support and coordinated care for physical comorbidities and social support \cite{skou2022multimorbidity,silverman2015american}. 
As a result, clinicians—\textcolor{black}{including psychiatrists, psychologists, pharmacists, nurses, and professionals offering occupational and social support in mental health services}—are forced to spend valuable time assembling disparate data points to create a coherent view for a patient's care, which can lead to inefficiency and delayed care \cite{hirsch2015harvest}.

Although electronic health records (EHRs) are designed to help clinicians manage information at the point of care, they often fail to present information in a format that offers effective cognitive support and mitigates information overload \cite{hirsch2015harvest}. 
This is largely because EHR systems mainly focus on recording information for individual patients and present data in tabular views or static text formats, with limited capability to highlight underlying trends in a patient's disease progression, reveal similarities within a team's caseload, or facilitate longitudinal health monitoring on a population scale \cite{badgeley2016ehdviz}. 
The need for better methods to manage and present increasingly complex information in EHRs has been long recognised \cite{west2015innovative,wang2022ehr}. Pioneering work in the 1990s introduced graphical summaries of test results and treatment data to enhance the presentation of patient status and reduce the burden of information overload \cite{powsner1994graphical}. 
Since then, research in this area has advanced by incorporating diverse datasets, including patient-generated data \cite{badgeley2016ehdviz,polhemus2022data} and knowledge bases \cite{shahar1999intelligent}, alongside emerging technologies, e.g. interactive visualisations \cite{plaisant2003lifelines}, statistical and machine learning analytics \cite{badgeley2016ehdviz}, to enhance the effectiveness and usability of visual analytics in healthcare.
These advancements have been applied across various domains, including patient data summarisation \cite{hirsch2015harvest,feblowitz2011summarization}, cohort search \cite{callahan2021ace}, care quality improvement \cite{elshehaly2020qualdash}, patient flow analysis \cite{soulakis2015visualizing}, population health management \cite{chishtie2022interactive}, and disease- or setting-specific pathway management  \cite{soulakis2015visualizing,huang2015novel,franklin2017dashboard}. 
Visual analytics has emerged as a powerful tool for converting complex health data into intuitive and visually-compelling presentations, enabling the extraction of valuable insights from big data and supporting informed clinical decision making \cite{caban2015visual}.

However, existing research has mainly focused on physical healthcare settings, with limited attention given to mental healthcare \cite{chung2020role}. 
Compared to physical health, visual analysis of mental health data presents greater complexity for two key reasons. 
First, mental health conditions often encompass both medical, psychological and social dimensions \cite{salvador2006framework}, requiring a broader range of data to achieve a comprehensive assessment of an individual's mental health. This highlights the need of an extensible toolkit that can facilitate the rapid design, development, and implementation of visual analytics across diverse data sources and types in mental healthcare. 
Second, treating mental health conditions typically involves a combination of interventions, including medication, psychotherapy, lifestyle changes, and social support services \cite{silverman2015american}. 
Unlike physical conditions, where structured data (such as blood assays and other characterizations) is more salient, \textcolor{black}{mental health assessments also rely on vast patient-reported information as quantitative measures to evaluate various aspects of mental health, including mental health presentations, relevant contextual factors, interventions, and outcomes \cite{roe2022patient}}. Such information is often documented in unstructured text, such as clinical notes and correspondences \cite{kadra2015extracting}. 
Integrating this unstructured information into visual analytics pipelines is more challenging compared to structured data such as numerical or categorical variables. \textcolor{black}{Thus, tools developed for physical health management in previous studies may not effectively process and represent mental healthcare data.}

Another notable limitation of many previous studies is their generalisability. 
First, most studies have focused on a specific clinical task \cite{hirsch2015harvest,callahan2021ace} or disease \cite{soulakis2015visualizing,ibrahim2020rapid}.
However, in \textcolor{black}{numerous} clinical settings, patients rarely present with a single disease or single risk factor that may affect their health \cite{schall2017usability}. 
There is a lack of integrated visual-analytics solutions that can systematically address the varied challenges faced by different healthcare partners, including clinicians and patients \cite{caban2015visual}. 
Second, existing solutions are often developed using proprietary or custom tools internally developed by hospitals or EHR system providers \cite{hirsch2015harvest,ibrahim2020rapid}, leading to limited interoperability with other tools and restricted applicability in different settings \cite{badgeley2016ehdviz}. 
Finally, for visual-analytics solutions to achieve better outcomes and effective adoption, they should be user-friendly, require minimal training, offer evidence-based recommendations, and integrate smoothly into clinicians' workflows \cite{schall2017usability}. 
Despite extensive focus on technical development in previous studies, there has been little effort to systematically develop and evaluate these solutions within the complex workflows of clinicians, nurses, and other health workers while considering their varied clinical priorities and patient needs \cite{lim2022toward}. 

In this work, we present VIEWER (Visual \& Interactive Engagement With Electronic Records), an open-source, cost-effective, and extensible toolkit created for the rapid design, development, and deployment of clinical information retrieval, analysis, and visualisation for supporting clinical decision making. 
VIEWER is an EHR-agnostic framework that utilises distributed information extraction pipelines, leveraging natural language processing (NLP) methods and open-source visualisation techniques to enable comprehensive search and visual analytics of a comprehensive health record from both structured and unstructured patient data within a health institution, rather than a curated dataset for a specific disease or patient cohort. 
We also systematically describe our interdisciplinary and collaborative approach to participatory design where we designed, implemented and evaluated VIEWER \textcolor{black}{on top of a bespoke EHR system within one of the largest National Health Services (NHS) Trusts for mental health in the United Kingdom (UK)}, aimed at addressing the diverse challenges faced by multiple healthcare partners. The partners included: (i) clinicians who need to synthesise disparate data to understand a patient’s condition within the context of their medical history, (ii) managers who require data-driven insights to optimise resource allocation and identify unmet needs, (iii) researchers who seek to understand disparities in outcomes across populations, and (iv) patients who wish to utilise their own medical data for self-monitoring. Our evaluation demonstrates the effectiveness of VIEWER in enhancing patient care in real-world clinical use cases and provides valuable insights into working collaboratively with clinical workers, researchers, patients and carers, and informaticians to iteratively refine and optimise informatics solutions for improved patient care.

\section*{Objective}

\textcolor{black}{This paper describes the development and implementation of VIEWER, an extensible visual-analytics system for improving clinical data accessibility and presentation for clinical decision support, and the system's proof-of-concept applications within mental health services. 
We first describe a participatory, iterative, and interdisciplinary collaborative design approach used to develop an initial prototype of VIEWER, implement it to extend the EHR ecosystem at the South London and Maudsley NHS Foundation Trust (SLaM), and continuously refine the ecosystem based on output from user feedback cycles.
We then detail the technical components of VIEWER, which includes distributed NLP-based information retrieval and extraction modules, and open-source visualisation services to provide visually interactive, problem-oriented, and holistic presentations of the patient record.
Finally, we conduct evaluations to assess VIEWER's utility and effectiveness in supporting healthcare professionals across various clinical workflows, such as patient health checks, medication reviews, and caseload management, thereby providing opportunities to identify areas for improvement of the implementation in its future iterations.}

\section{Materials and Methods}

\subsection{Setting}
The VIEWER platform was developed within the National Institute for Health and Care Research (NIHR) Maudsley Biomedical Research Centre (BRC) and has been deployed at SLaM, the UK’s largest mental health NHS Trust. SLaM provides more than 240 secondary mental health services\footnote{In the UK, secondary mental health services refer to mental health services in secondary care, typically including hospitals, some psychological wellbeing services, community mental health teams (CMHTs), crisis resolution and home treatment teams (CRHTs), assertive outreach teams and early intervention teams.} for a local population of 1.3 million residents in South London. In addition, SLaM provides more than 50 national and specialist services for people across the UK and beyond. Each year, the Trust provides inpatient care for over 5,000 people and treat more than 40,000 patients.

As one of the earliest adopter of an EHR system in the UK, SLaM has used its bespoke EHR system called Electronic Patient Journey System (ePJS) for a centralised management of patient data since 2006. This system stores a comprehensive record of all clinical information recorded throughout a patient's journey when accessing services from the Trust, including demographic and contact information, details of referrals and transfers, clinical assessments, care plans and medications, clinical activity and reviews, as well as correspondence letters and supporting care notes between healthcare professionals. The record consists of both structured data (e.g., numerals, dates and selection-lists) and unstructured free text (e.g., written assessments, progress notes and correspondence). 

Like other EHR systems \cite{hirsch2015harvest}, ePJS is a Web-based application designed for entering and reviewing patient information, with a specific focus on displaying plain textual data rather than visual graphics. Patient summaries within a clinician's caseload are presented in a tabular format, where each row provides a link to a patient's profile. This profile contains various tabs corresponding to different components of patient details, including clinical notes, laboratory tests, diagnoses, and medication prescriptions. All data from the clinical record are fine-grained information at the individual patient level, with no aggregation at the team/population levels or aggregation of temporal information for the same patient. A basic search function allows users to search based on document type, test type, recording date, patient name, and their unique identifiers such as the national NHS number or local Trust identifier (patient IDs). However, the clinical data in free-text documents, constituting the majority of the EHR, are less amenable for search queries and less accessible to clinicians.

The NIHR Maudsley BRC, the UK’s BRC focused on mental health and hosted by SLaM, has a long-standing tradition of clinical research using EHR data. A key infrastructure supporting this research is the Clinical Records Interactive Search (CRIS), a case register and governance model that allows researchers to use de-identified EHR data for various research purposes, including epidemiology, health data linkage, and clinical natural language processing (NLP) \cite{stewart2009south}. Additionally, the BRC developed CogStack, an open-source platform for clinical data retrieval and extraction that enables semantic search of free-text data, risk alerting, and data visualisation to support clinical decision-making \cite{jackson-2018-cogstack}. VIEWER builds on the aforementioned research infrastructure by incorporating additional capabilities from clinically-focused user feedback, thereby facilitating large-scale use in clinical practice.


\begin{table*}
\scriptsize
\caption{Needs identified for different partners.}
	\label{tab:needs}
\begin{tabular}{l|l}
\hline
Role                                           & Needs                                                                                         \\                                            \hline
\multirow{4}{*}{Service director}              & Detect specific needs within local populations                                                \\
                                               & Identify how well these needs are being met                                                   \\
                                               & Monitor trends in data over time                                                            \\
                                               & Identify inequalities in access                                                              \\                                             \hline
\multirow{5}{*}{Team manager}                  & Ensure workloads are being distributed evenly                                                 \\
                                               & Identify specific needs in the caseload and how well these needs are being met                \\
                                               & Identify trends over time to support improvement work                                         \\
                                               & Have a ``single version of the truth'' to support transitions of care                             \\
                                               & Identify patients on different stages of their care with the team                             \\                                            \hline
\multirow{3}{*}{Clinical worker} & Identify patients who may benefit from their input                                            \\
                                               & Track patient parameters and outcomes to evaluate effectiveness/tolerability of interventions \\
                                               & Deliver evidence-based interventions based on clinical needs and effectiveness                \\                                            \hline
\multirow{4}{*}{Care coordinator}             & Ensure equitable division of time based on needs                                               \\
                                               & Deliver proactive care based on identified needs                                              \\
                                               & Ensure continuity of care                                                                     \\
                                               & Track patient parameters and outcomes to evaluate effectiveness/tolerability of interventions \\                                            \hline
\multirow{3}{*}{Patient and carer}             & Cross-check service data against their own understanding                                      \\
                                               & Track their data over time                                   \\
                                               & Track key metrics for their care against good practice  \\ 
                                            \hline
\end{tabular}
\end{table*}

\subsection{\textcolor{black}{Interdisciplinary collaboration}}
VIEWER was initiated by an interdisciplinary team, which consisted of two clinicians, two computer scientists, and two health informaticians, at SLaM in January 2020, aiming to enhance data-driven clinical decision support and promote population health management within the Trust. 
To ensure that the perspectives, needs, and preferences of all partners are considered, an iterative participatory design approach \cite{jensen2018bridging,sedlmair2012design}, a widely used approach for clinical decision support systems \cite{clemensen2017participatory,jensen2018bridging}, was adopted throughout the design, development, and deployment stages. This approach involves a collaborative process, whereby all partners engage with each other to co-create and co-improve the development and deployment, with the aim of developing solutions that are usable, useful, and effective.

Our design process included three phases. 
In phase 1, partner needs were identified and discussed. 
These needs stem from 1) brainstorming sessions within the development team, and 2) feedback obtained from prototype demo sessions and design workshops involving the various partners, including clinical workers.
\textcolor{black}{Table \ref{tab:needs} summarises the resulting needs identified from different categories of partners in this phase.} 
Phase 2 involved the design and development of a series of prototypes, leveraging interdisciplinary skills from the development team. The team held routine weekly meetings to identify and prioritise areas for improvement, brainstorm practical solutions, and plan the next iteration of development. Each prototype underwent testing, \textcolor{black}{which involves: 1) validating the accuracy of information presented in VIEWER by comparing it with the raw data in ePJS, particularly for data processed by NLP pipelines such as medication extraction \cite{wang2023unraveling}, 2) assessing the appropriateness of the information's visualization, including the suitability of visual graph types and labeling, and 3) the accessibility of the visual interface, focusing on logical design, layout coherence, and integration with other system components.
A prototype} was eventually migrated into a production environment after reaching consensus by all relevant partners.
\textcolor{black}{Key results from this phase are showcased through the user interfaces detailed in Section \ref{section_interface}.}
In phase 3, each new prototype was piloted and evaluated in a real clinical setting. During this phase, demos and training materials, including user guides, screenshots, and recorded video tutorials, were provided to pilot teams. 
\textcolor{black}{The pilot sessions were centered on specific use cases and involved participants with diverse backgrounds, demographics, and areas of expertise. Participation was voluntary from teams that requested involvement or expressed interest in a given use case. No formal selection process was implemented, ensuring organic representation to promote diversity, inclusivity, and equity in engagement.}
Feedback was actively collected during demos, training sessions, and feedback channels (e.g. emails and review sessions) to guide the next iteration of development. 
\textcolor{black}{The key outcomes from this phase are outlined in the use cases in Section \ref{section_use_cases} and the system usability findings in Section \ref{section_engage}.}

\textcolor{black}{It is important to highlight that the design process followed an iterative approach. Requests from partners and feedback from end-users were continuously gathered and prioritized by the development team to refine a current version and identify areas for improvement in subsequent versions. This process facilitated ongoing feedback loops, which ultimately led to the creation of a final design that more closely aligned with the users' needs and expectations.}



\begin{figure*}
	\includegraphics[width=1\columnwidth]{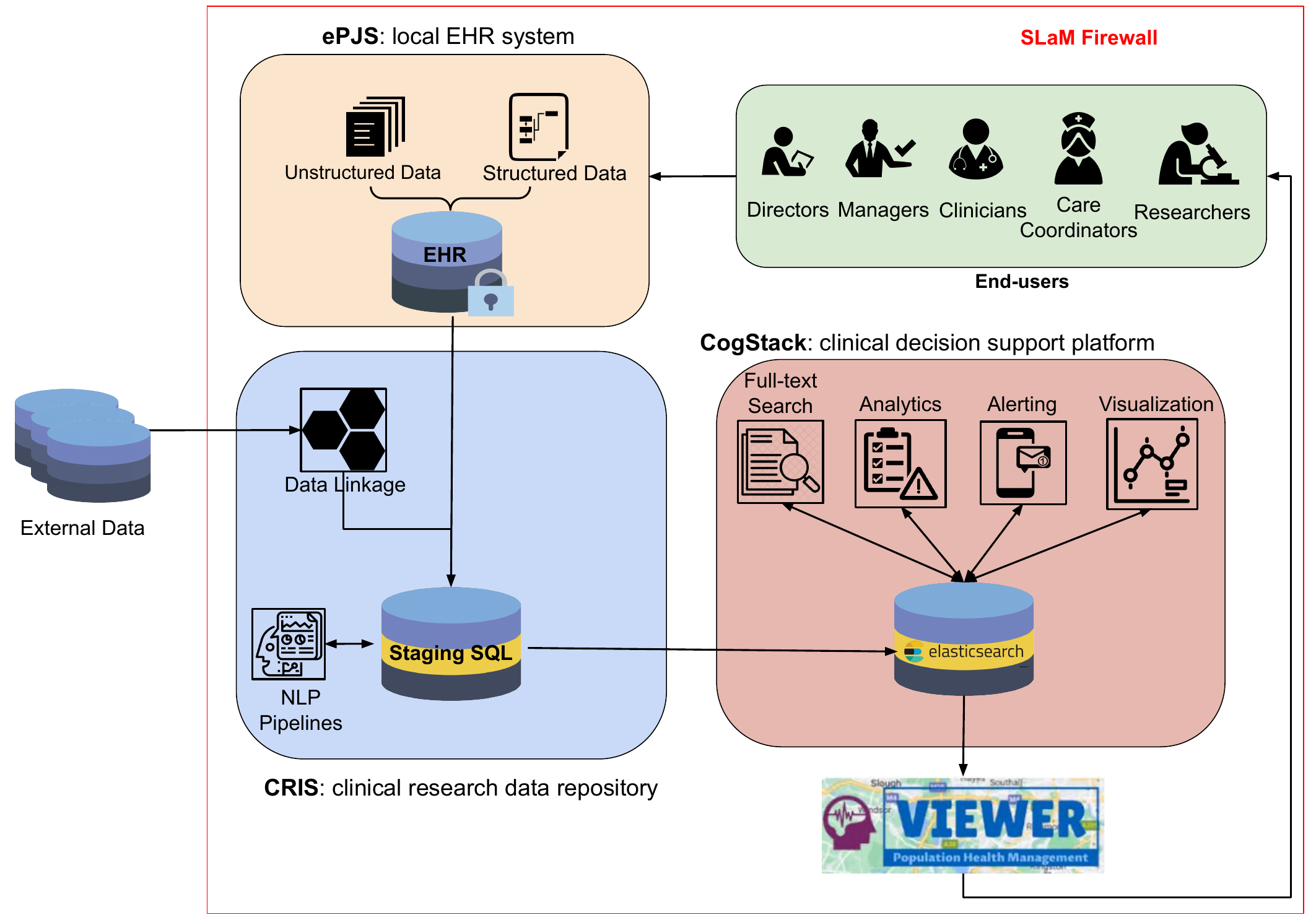}
    \caption{VIEWER system architecture. VIEWER is a Web-based application built on three pre-existing systems. 1) ePJS: the EHR system of SLaM; 2) CRIS: a case register and clinical research data repository used to develop a data model by leveraging its data linkage and NLP pipelines; 3) CogStack: an open-source data retrieval and extraction platform used to inform clinical decision-making by semantic search of free text, data analytics, visualisation and risk alerting. All data and pipelines are held securely within the Trust firewall on Azure cloud servers.}
    \label{fig:architecture}
\end{figure*}

\subsection{Overall architecture}
Figure~\ref{fig:architecture} illustrates the architecture and pipelines used in VIEWER. Patient data, comprising both structured and unstructured information, is initially collected and recorded in a SQL database using ePJS, SLaM's EHR system, at the point of care. 
The updated EHR in ePJS is synchronised daily with a staging SQL database in the CRIS system using Extract, Transform, and Load (ETL) scripts. 
This staging environment allows the ETL processes to not only handle raw data in ePJS, but also to integrate NLP pipelines for extracting key medical information (e.g. symptoms, interventions, outcomes, and contextual factors) from free-text notes \cite{noauthor_cris_nodate}, and data linkage connecting local data with other clinical data sources to enhance the richness of EHR \cite{noauthor_cris_nodate-1}. 
\textcolor{black}{The NLP pipelines extract mentions of medical entities, such as symptoms and interventions, from clinical notes using rule-based and machine learning approaches built on the General Architecture for Text Engineering (GATE) software. GATE is a comprehensive suite of tools used for diverse NLP tasks such as text parser, morphology, tagging and information extraction \cite{cunningham2013getting}. 
These NLP pipelines have been regularly validated (with precision and recall metrics reported and updated in \cite{noauthor_cris_nodate}) and deployed to extract clinical information from routinely collected EHR in CRIS. 
Details on the development and validation of the NLP applications within CRIS are documented in previous publications \cite{kadra2015extracting, perera2016cohort, cunningham2013getting} and online resources \cite{noauthor_cris_nodate}.}
We designed and tested the ETL pipelines based on de-identifiable data in CRIS and plug in these pipelines to raw EHR from ePJS to enable clinical use of data.

The harmonised data model is then ingested into CogStack using customised Python scripts with Elasticsearch Application Programming Interface (APIs) and Apache NiFi\footnote{\url{https://github.com/CogStack/CogStack-NiFi}}. 
Interactive visualisations of the ingested data and dashboards are created using the Kibana component in CogStack. 
We used the open-source versions of Elasticsearch and Kibana from Open Distro\footnote{\url{https://opendistro.github.io/for-elasticsearch-docs/}} in our implementation. 
To enhance user experience, a lightweight, customised wrapper application is created, serving as interfaces that logically organise individual dashboards based on clinical use cases. End-users can seamlessly navigate from VIEWER to ePJS to take action (e.g. adjusting care plan for a patient) based on insights provided by VIEWER. User authentication and access control are handled by the EHR system using the Lightweight Directory Access Protocol (LDAP) \cite{sso}. 
For more technical details of each component in the architecture, see our previous papers \cite{stewart2009south,jackson-2018-cogstack}.

\begin{figure*}
	\includegraphics[width=1\columnwidth]{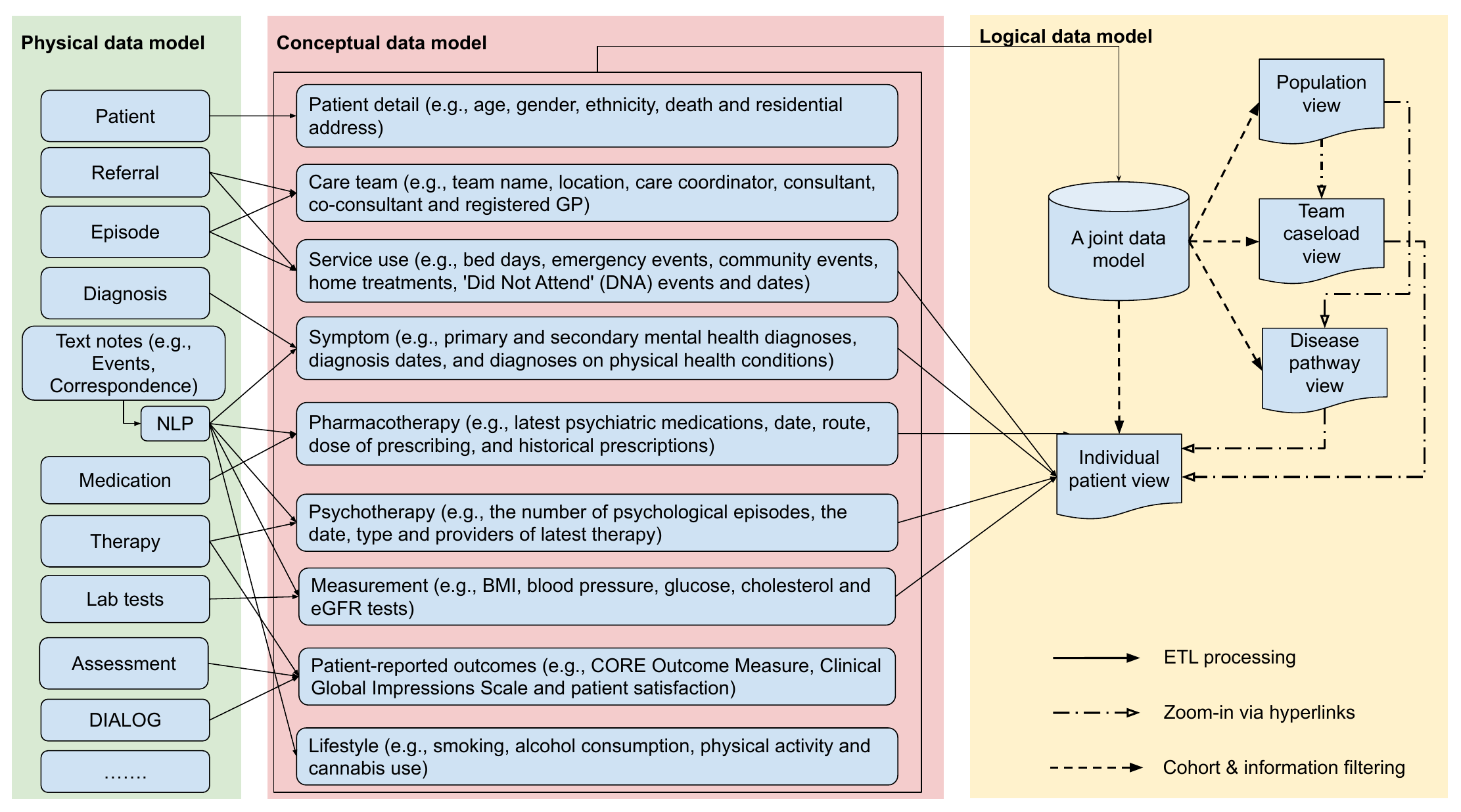}
    \caption{VIEWER data models. The physical data model illustrates physical data stores of the source EHR data, the conceptual data model describes the conceptual structure of entity classes and the logical data model describes the semantic and logical relationships among entities.}
    \label{fig:data}
\end{figure*}

\subsection{Data model}
To meet the varying needs of different partners, we have crafted a comprehensive data model, encompassing a diverse array of patient attributes, clinical activities, and outcomes. As shown in Figure \ref{fig:data}, we first identify relevant data sources within the source EHR, including both structured information and unstructured text data. These sources are then transformed and conceptualised into entity classes based on local settings and project-specific requirements. 

The conceptual entities are then joined into a unified data table to consolidate the most recent information for each service user in the Trust. This consolidated data model undergoes daily updates and offers analytical and visual capabilities at various levels, ranging from a holistic population view to team-level caseload perspectives, disease-level pathway insights, and individual patient profiles. All these views are interconnected, allowing seamless navigation through embedded hyperlinks.   
For a granular patient-level examination, longitudinal data for time-dependent entities, such as service utilisation, symptoms, treatments, and measurements, are incorporated, enabling the monitoring of patients' health statuses and care plans over time.

\subsection{Interface design}
The front-end interface of VIEWER consists of Web-based visualisations and dashboards developed using Kibana, an open-source data visualisation tool available in CogStack \cite{jackson-2018-cogstack}. This contains four categories of dashboards, including ``Population Health'', ``Clinical Pathways'', ``Caseload Management'', and ``Patient Chart''.
Given the stringent safety requirements, time-sensitive nature of clinical practice, and varying technical proficiency among end-users, we prioritise the following principles in the visualisation design:
\begin{itemize}
    \item \textbf{Clarity}: Ensure that visualisations remain clear, simple, and user-friendly, avoiding unnecessary complexity that could confuse end-users or increase training costs.
    
    \item \textbf{Accuracy}: Employ accurate graphs, labels and representations to ensure precise representation of data in visualisations and facilitate cross-checks of information across different visualisations (e.g. providing source text snippets for NLP extraction).
    
    \item \textbf{Relevance}: Only include relevant information that supports the intended message in the visualisations to avoid information overload.
    
    \item \textbf{Integration}: Ensure that the interface and content of visualisations are seamlessly aligned with clinical workflows and integrate seamlessly with EHR systems. Individual components are interconnected as a cohesive, self-contained system, e.g. enabling flawless navigation from population-level visualisations to caseload and individual patient views.
    
    \item \textbf{Interactivity}: Add interactive features (e.g. filtering and searching) when appropriate to allow users to explore the data further, gaining deeper insights based on their interests, beyond pre-defined information in default interfaces.   

    \textcolor{black}{\item \textbf{Learnability and error recovery}: To enhance learnability, employ consistent design elements such as icons, buttons, terminology, and structured layouts, while avoiding technical jargon. Additionally, incorporate default interfaces with ``undo'' options and accessible help guides to facilitate error recovery and ensure users can easily correct mistakes.}
    
\end{itemize}

These dashboards are organised in a logical manner through a lightweight, configurable Web-based wrapper based on the Java Spring framework. 
When an authenticated user selects a dashboard via the navigation page (\textit{SI} Figure S2), data are queried in Elasticsearch based on pre-defined filters in the dashboard and the data are initially visualised in default formats. Users can interact with the interface by employing features like filtering, zooming, dragging, drilling down, selecting time ranges, and conducting full-text searches and queries to explore data and derive insights based on their interests. Smooth transition across dashboards for additional information and navigation from VIEWER to the source EHR system for actions are facilitated by hyperlinks based on Kibana scripted fields. Each visualisation and dashboard is accompanied by short descriptions and banners that provide essential information about the data and visualisations.

Two versions of visualisations are available based on the use cases: the ``clinical'' version, which includes identifiable patient data for direct clinical tasks such as patient review and caseload management, and the ``non-clinical'' version, where patient identities are anonymised for non-clinical tasks like population based resource allocation, quality improvement and demos/presentations. The de-identified data in the non-clinical version is generated using Kibana's scripted fields and field anonymisation features.

\subsection{Scalability} 
To ensure scalability of the system, we created a distributed computing infrastructure using Docker Swarm. 
Since all the data queried by end-users, search engine, and front-end interfaces are within CogStack, our focus is on enhancing the scalability of the CogStack instance. 
See supplementary information (\textit{SI}) Figure S1 for our distributed deployment of CogStack based on Docker Swarm on Azure. 
Processed data, extracted entities, audit logs, and metadata are stored in a distributed Elasticsearch cluster spanning three nodes, and hence new data are added and integrated seamlessly without system downtime.
The distributed nature of this deployment also enables us efficiently manage and scale the system to accommodate a growing user base, increasing data volumes, and additional computational tasks over time.

\subsection{Data security and audit}
This study was approved by multiple information governance bodies at SLaM, including the CRIS Oversight Committee for the development of the data model using de-identified data, the CogStack Oversight Committee for the VIEWER prototype utilizing identifiable data, 
where Secondary analysis of CRIS data was approved by Oxford Research Ethics Committee C (reference 18/SC/0372).
Data Protection Impact Assessment and Clinical Risk Assessment have been conducted to ensure the development and deployment of VIEWER complies with the Trust's data security and patient safety policies.


All data, infrastructure, and software used in VIEWER are securely housed within the Trust's firewall and are accessible only to authorised users.
Two user roles have been established to govern access to the front-end interfaces based on index-, document- and field-level access control within the Kibana security plugins. 
A clinical role, providing access to identifiable patient data, is assigned to clinical staff who already have access to such data in the Trust's EHR system. Conversely, a non-clinical role, providing access solely to the de-identified version, is assigned to non-clinical staff, such as managers and authorised researchers. 
Users with a clinical role can access both ``clinical'' and ``non-clinical'' versions of data, while users with a non-clinical role can only access ``non-clinical'' version.
User activities, including logins and queries, are meticulously logged using Kibana audit plugins for audit purposes, ensuring that only relevant clinical information is accessible for a specific user.

\section{Results}

\subsection{Patient characteristics}
As of a recent census date on March 25th 2024, VIEWER ingested information for all patients (n=409,870) who had a team episode or ward stay at SLaM, where 51,057 were active caseload, namely those currently managed by the Trust. \textcolor{black}{Table \ref{tab.patient} summarises the patient characteristics for both active and inactive/discharged caseloads that have been integrated into VIEWER}. Compared to the inactive caseload, patients in the active caseload tend to younger and have a higher proportion from the Black and Mixed ethnicity groups. Also, active caseload has a higher prevalence of severe mental health conditions, such as schizophrenia (ICD 10 code F2), bipolar disorder (F31) and major depressive disorder, and a higher proportion of living in the SLaM catchment area.

\begin{table*}
\scriptsize
\centering
\caption{Descriptive statistics of patient characteristics, grouped by active and inactive caseloads. Mann–Whitney U test is used to assess the statistical significance of differences between groups for numerical variables and the Chi-square test is used for categorical variables. Adjusted p-values with the Bonferroni correction were reported. }
\begin{tabular}{p{8.5cm}llllll}
\hline
           & All             & Active         & Inactive        &     $s$     &  $p$ \\
\hline
\hspace*{-1em} Number of patients (\%)           & 411,313 (100)        & 51,173 (12.44) & 360,140 (87.56) &          &   \\
\hspace*{-1em} Age        & 40.30 (18.55)   & 33.34 (19.11)  & 41.29 (18.25)   & 6.83E+09 & <0.001 \\

\hspace*{-1em} Gender &                 &                &                 & 144.87   & <0.001 \\
  Female     & 212,244 (51.60) & 26,869 (52.51) & 185,375 (51.47) &          &   \\
  Male       & 198,068 (48.16) & 24,065 (47.03) & 174,003 (48.32) &          &   \\
  Unknown & 1,001 (0.24)    & 239 (0.47)     & 762 (0.21)      &          &   \\

\hspace*{-1em}Ethnicity  &                 &                &                 & 12226.57 & <0.001 \\
White      & 172,042 (41.83) & 23,243 (45.42) & 148,799 (41.32) &          &   \\
Black      & 58,977 (14.34)  & 10,674 (20.86) & 48,303 (13.41)  &          &   \\
Asian & 20,129 (4.89)   & 2,783 (5.44)   & 17,346 (4.82)   &          &   \\
Mixed      & 17,183 (4.18)   & 5,062 (9.89)   & 12,121 (3.37)   &          &   \\
Other & 23,368 (5.68)   & 3,139 (6.13)   & 20,229 (5.62)   &          &   \\
Unknown    & 119,614 (29.08) & 6,272 (12.26)  & 113,342 (31.47) &          &   \\

\hspace*{-1em}Diagnosis  &                 &                &                 & 9828.9   & <0.001 \\
F0 - Organic, including symptomatic, mental disorders         & 14,938 (3.63)   & 1,256 (2.45)   & 13,682 (3.80)   &          &   \\
F1 - Mental and behavioural disorders due to psychoactive substance use        & 31,621 (7.69)   & 3,613 (7.06)   & 28,008 (7.78)   &          &   \\
F2 - Schizophrenia, schizotypal and delusional disorders & 19,415 (4.72)   & 5,528 (10.80)  & 13,887 (3.86)   &          &   \\
F3 - Mood [affective] disorders         & 42,123 (10.24)  & 3,440 (6.72)   & 38,683 (10.74)  &          &   \\
F4 - Neurotic, stress-related and somatoform disorders        & 44,523 (10.82)  & 4,380 (8.56)   & 40,143 (11.15)  &          &   \\
F5 - Behavioural syndromes associated with physiological disturbances and physical factors         & 13,376 (3.25)   & 1,557 (3.04)   & 11,819 (3.28)   &          &   \\
F6 - Disorders of adult personality and behavior         & 9,189 (2.23)    & 1,531 (2.99)   & 7,658 (2.13)    &          &   \\
F7 - Mental retardation        & 2,675 (0.65)    & 412 (0.81)     & 2,263 (0.63)    &          &   \\
F8 - Disorders of psychological development        & 12,225 (2.97)   & 2,226 (4.35)   & 9,999 (2.78)    &          &   \\
F9 - Behavioural and emotional disorders with onset usually occurring in childhood and adolescence        & 23,622 (5.74)   & 3,768 (7.36)   & 19,854 (5.51)   &          &   \\
F99 - Unspecified mental disorder       & 38,000 (9.24)   & 1,827 (3.57)   & 36,173 (10.04)  &          &   \\
FX - No Axis 1 disorder        & 19,969 (4.85)   & 2,032 (3.97)   & 17,937 (4.98)   &          &   \\
Z - Factors influencing health status and contact with health services & 44,533 (10.83)  & 4,480 (8.75)   & 40,053 (11.12)  &          &   \\
Other illness        & 6,391 (1.55)    & 246 (0.48)     & 6,145 (1.71)    &          &   \\

\hspace*{-1em} Borough of residence    &                 &                &                 & 2780.83  & <0.001 \\
Croydon    & 71,815 (17.46)  & 10,929 (21.36) & 60,886 (16.91)  &          &   \\
Lambeth    & 66,144 (16.08)  & 10,498 (20.51) & 55,646 (15.45)  &          &   \\
Lewisham   & 62,647 (15.23)  & 8,625 (16.85)  & 54,022 (15.00)  &          &   \\
Southwark  & 62,434 (15.18)  & 7,430 (14.52)  & 55,004 (15.27)  &          &   \\
Homeless   & 6,521 (1.59)    & 438 (0.86)     & 6,083 (1.69)    &          &  \\
Other & 141,752 (34.46) & 13,253 (25.90) & 128,499 (35.68) &          &   \\
\hline
\end{tabular}

\label{tab.patient}
\end{table*}

\subsection{Data characteristics}
The VIEWER platforms efficiently ingest more than 1.7 million documents daily from source EHR systems, and support information retrieval and visualisations across more than 551 million documents for more than 1,000 users across the Trust. Detailed characteristics of the data used in VIEWER are summarised in \textit{SI} Table S1. Most time series data of patients were used to visualise patient charts, enabling clinicians to investigate longitudinal information for individual patients. In contrast, most visualisations for population, caseload, and pathway management are based on the information in the ``Full caseload'' table, as it reflects the most recent and comprehensive details of a patient's care. The biggest table ``Full caseload snapshots'' contains time series of the ``Full caseload'' table, allowing the Trust or a team to monitor their service delivery over time.

\subsection{User interfaces} 
\label{section_interface}

\begin{figure*}
	\includegraphics[width=1\columnwidth]{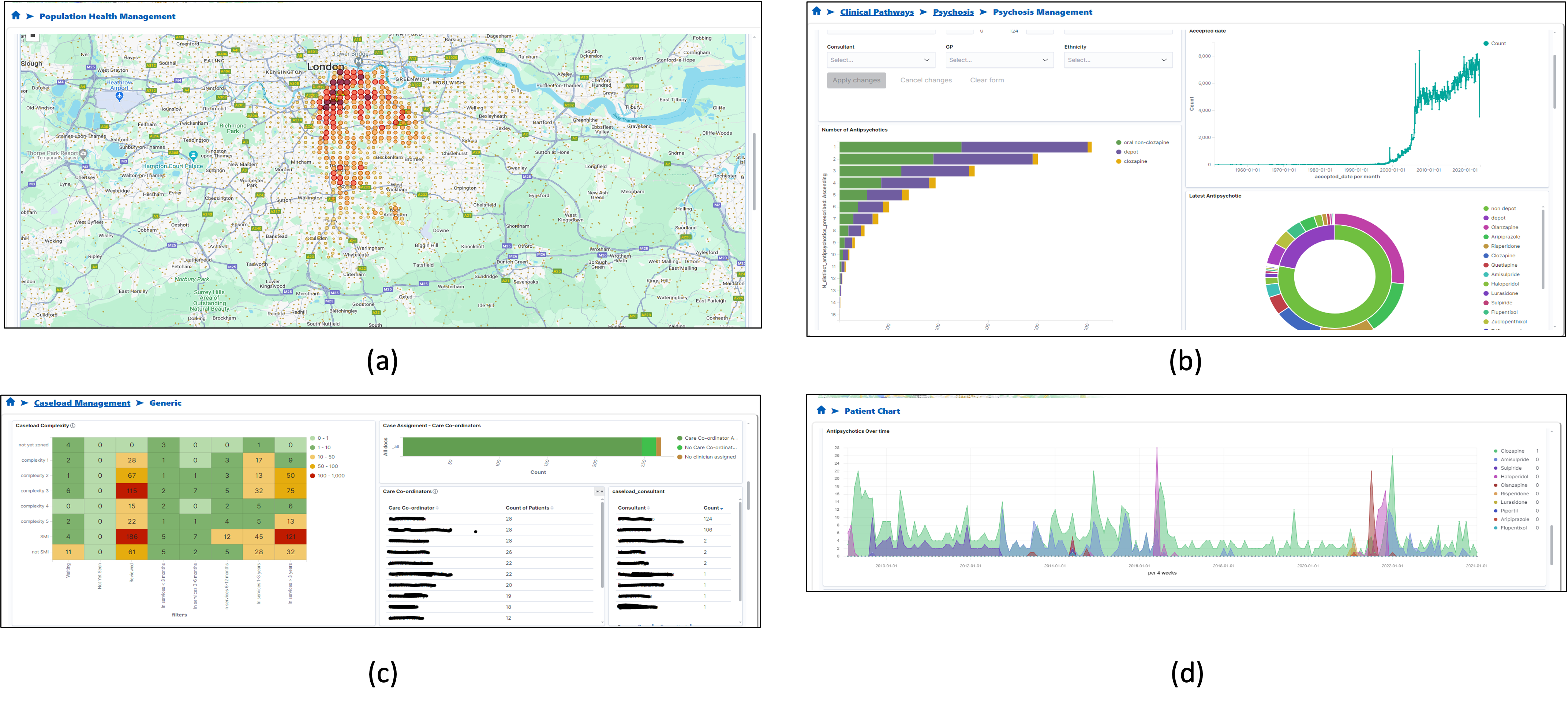}
    \caption{Screenshots for different dashboards in VIEWER. (a) Maps of residences for all active service users of the Trust in the ``Population Health'' view; (b) Patterns of anti-psychotic medication use in patients with psychosis in the ``Clinical Pathways'' view; (c) Case complexity in a care team's caseload, stratified by duration of service use, care coordinators and consultants in the ``Caseload Management'' view. and (d) Anti-psychotic prescriptions over time for a patient in the ``Patient Chart'' view.}
    \label{fig:screen}
\end{figure*}

As shown in Figure \ref{fig:screen}, VIEWER provides web-based interactive, problem-oriented visualisations of patient data through multiple perspectives, including: 
\begin{itemize}[leftmargin=*]
    \item \textbf{Population health}: This category of visualisations aim to offer a comprehensive overview of the whole population served by the Trust, encompassing demographics, geographical distribution, diagnoses, outcomes, and care teams. It provides multidimensional filters that allow users to delineate sub-populations and discern specific needs, such as identifying regions with elevated incidents of new cases, those with stretched resources and inequalities in service access. 

    \item \textbf{Clinical pathways}: This category includes a set of dashboards designed to facilitate proactive care through a multidisciplinary approach to the assessment, diagnosis, and treatment of individuals experiencing symptoms associated with specific mental disorders, such as psychosis and bipolar disorders. Visualisation components within this category include identifying cohorts of patients who have not received interventions and may benefit from them based on historical treatments and responses. It also allows to identify patients who need a medication review based on their side effect profiles and those showing early signs of relapse or non-adherence to treatments, and assess who has met national care standards, including prescribing, psychological therapy, and physical health. 
    For discharged patients, it facilitates coordination with their primary care physician to anticipate and prevent relapse, as well as reduce reliance on crisis pathways. Additionally, it provides an overview of the entire cohort with a specific diagnosis, including both those currently under care and those who have been discharged.
    
    \item \textbf{Caseload management}: This category focuses on enhancing caseload management within a clinical team. It involves visualizing uneven distributions of case risk and complexity across team members, identifying factors contributing to risk/complexity, highlighting potential indicators of unmet needs (such as crisis care, patients' clinical outcomes, and their satisfaction with services), monitoring care continuity during leaves or absences. This also enables the MDT within the team and the broader system to identify where limited resources should be allocated to achieve better outcomes.

    \item \textbf{Patient chart}: This category encompasses visualisations that provide detailed, individual-level information, offering insights into the current state of a specific patient (e.g. recent interventions and outcomes) as well as historical data on interventions and outcomes over time. By integrating longitudinal data from multiple entities, clinicians can efficiently comprehend the overall care trajectory of a given patient. This facilitates cross-checks of care data against clinicians' own understanding and enables the identification of personalised needs and care plans for individual patients. 

\end{itemize}

\subsection{Clinical applications}
\label{section_use_cases}

\begin{figure*}
	\includegraphics[width=1\columnwidth]{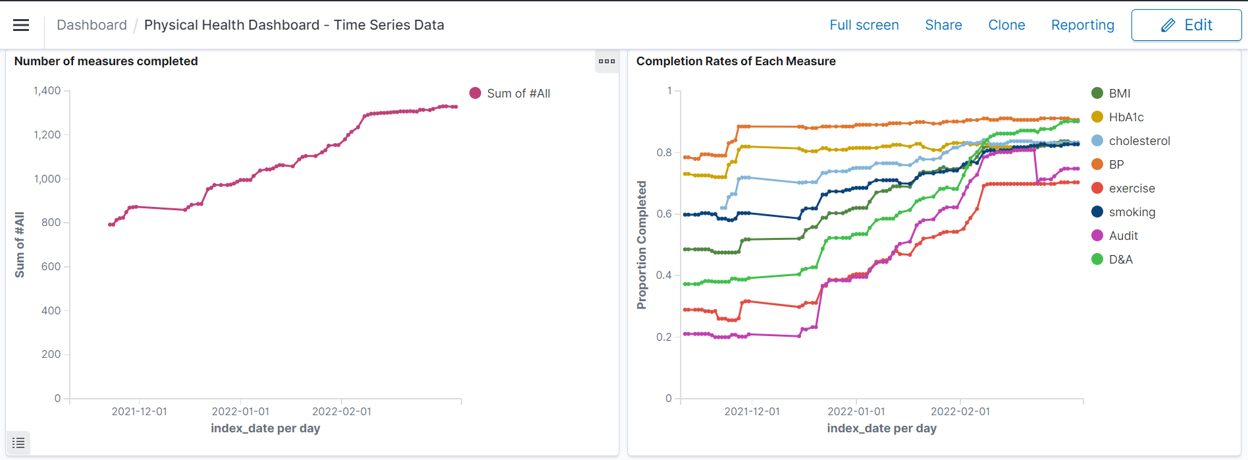}
    \caption{The number of measures completed and completion rate of each measure in annual physical health check over time for patients managed within a team.}
    \label{fig:physical}
\end{figure*}

\subsubsection{Physical health monitoring}
The first use case of VIEWER was to support annual physical health check \textcolor{black}{for all active patients within the Trust, particularly for those living with severe mental health conditions or undergoing long-term psychiatric medication treatment}. 
Compared to the general population, people with mental health conditions often face poorer physical health \cite{thornicroft2011physical}, which has contributed to reduced quality of life and a shortened life expectancy by 10-20 years for people with severe mental health conditions \cite{john2018premature,chesney2014risks}. 
\textcolor{black}{This disparity is partly due to the potential side effects of psychiatric medications, such as weight gain, metabolic syndrome, cardiovascular risks, and other chronic conditions \cite{smith2013schizophrenia,wang2022patient}}.
To enhance overall health and decrease premature mortality among this group, clinical guidelines recommend annual physical health checks to promptly identify and manage physical health conditions\footnote{\url{https://www.england.nhs.uk/long-read/improving-the-physical-health-of-people-living-with-severe-mental-illness/}}. 
These assessments typically include the evaluation of multiple parameters, including blood pressure, body mass index (BMI), blood glucose or HbA1c levels, lipid profile, smoking status, alcohol consumption, and lifestyle factors like diet and physical activity levels, which can be time consuming.

VIEWER has served as the primary system supporting a team of six members in overseeing physical health checks for the entire active caseload of over 51,000 patients within the Trust. VIEWER provided a comprehensive overview of physical health checks by leveraging NLP-based information extraction methods. Through interactive visualisations, team members can readily identify patients who require a physical health assessment. Each identified patient is provided with a hyperlink that directs clinicians to their profile within the EHR system, facilitating actions such as scheduling a check or contacting GPs for additional information. After completion, these checks are recorded in the source EHR system and then incorporated into VIEWER.
This systematic approach has streamlined the check process and resulted in an increase in completed assessments. For example, Figure \ref{fig:physical} shows the rise in the number of measures completed by the team after adopting VIEWER as a systematic approach within a short period of two months.

\subsubsection{Medication review}
VIEWER's second use case was to facilitate medication review \textcolor{black}{for 164,819 patients with documented prescriptions for psychiatric medications, including antipsychotics, antidepressants, and mood stabilisers}.
Many individuals in secondary mental health services take psychiatric medications, often for long periods \cite{moncrieff2016results}. 
Regular reviews are essential for optimizing treatment outcomes, improving adherence to prescribed medications, and minimizing adverse effects \cite{kane2013non}. 
This process requires a thorough assessment of all medication records within the entire EHR. Since most prescribing information and related contextual details (e.g. adverse reactions and non-adherence) are recorded in unstructured text such as clinical notes and correspondences between healthcare professionals (e.g. mental healthcare specialists, pharmacists, and GPs), each review typically takes hours to complete in the local Trust.

VIEWER enables to reduce the review time through user-friendly visualisations of a comprehensive medication data model, which integrates medication information extracted from both structured data and unstructured text using NLP techniques. 
In a preliminary evaluation of this new method, a pharmacist reviewed three medication cases previously completed by another pharmacist. For each case, a comparison was made between the medications identified by the human pharmacist and those generated by VIEWER. The results showed that 100\% (n=20) of the antipsychotics, antidepressants, and mood stabilisers identified in the human reviews (n=3) were also present in the VIEWER-generated lists. VIEWER also identified four extra medications not mentioned in the human reviews. Of these, 75\% (n=3) were treatments that had been considered but not started, and 25\% (n=1) was an antidepressant that had been trialed. Furthermore, the time spent per review using VIEWER was reduced from 1-2 hours to 10-20 minutes.

\subsubsection{Caseload management}

The last use case of VIEWER discussed here focuses on enhancing caseload management to meet the requirements set by the Care Quality Commission (CQC), an independent regulator of health and adult social care in England. Caseload management involves various aspects of a patient’s mental health needs ranging from medication management, psychotherapy, social support to resource allocation. Effective caseload management enables healthcare providers to improve efficiency in resource use, reduce waste and maximise patient outcomes. Table \ref{tab:caseload} lists exemplar tasks supported by VIEWER in the local mental health Trust. Since the Trust's EHR system was primarily developed for recording clinical data at the point of care for individual patients, it has limited capability for retrieving information for a group of patients. These tasks thus often require intensive manual work to review each patient's records and aggregate the data. This process is not only time-consuming but also introduces inconsistencies in data collection, analysis and recording. 

By integrating data aggregation processes in data ingestion pipelines and filters within Kibana visualisations, VIEWER allows team members to \textcolor{black}{access consistent and unified information about their caseloads}. Also, interactive data visualisations in VIEWER enable users to complete these tasks with just a few clicks, significantly improving efficiency. For example, team leaders need to regularly review the number of patients under each care coordinator during supervision, a task that typically takes at least 20 minutes per review. In contrast, this can be accomplished with a single click of a filter for a team name or a care coordinator's name in VIEWER. Additionally, VIEWER's rapid development capability has enabled the Trust to meet CQC requirements and deliver results for an annual inspection within just 6 months.

\begin{table*}
\scriptsize
\centering
\caption{Caseload management tasks supported by VIEWER.}
\begin{tabular}{p{3.5cm}p{5cm}p{6.2cm}}
\hline
Task & Purpose    & Examples     \\ \hline
\multirow{7}{3.5cm}{Prioritisation and Plan}    & \multirow{7}{5cm}{Evaluating the needs of each patient and prioritizing care based on the severity and urgency of their conditions}                                           & - Identify Red-Zone patients. \\ 
&& - Locate accident and emergency (A\&E) attendances. \\ 
&& - Assess complexity and risk factors for a patient.\\
&& - Plan outreach care such as home visits based on geography.                      \\ \hline
\multirow{8}{3.5cm}{Resource Allocation}        & \multirow{8}{5cm}{Effectively allocating resources, including staffing and services, to ensure that all patients receive the appropriate level of care.}                                                & - Review caseload for each care coordinator.  \\ 
&& - Allocate new referrals and assign caseload across care coordinators based on complexity, geography/GP catchment area, rather than number. \\ 
&& - Detect unassigned patients. \\ 
&&- Identify patients eligible for earlier discharge based on complexity, risk and use of crisis services. \\ 
\hline

\multirow{14}{3.5cm}{Coordination of Care}       & \multirow{14}{5cm}{Ensuring that patients receive comprehensive care by coordinating services among various healthcare providers, such as psychiatrists, therapists, primary care physicians, and social workers.} & - Facilitate multi-disciplinary team (MDT) in-reach opportunities to support care coordination by psychology, physical health, occupational therapy social care etc., based on DIALOG outcomes and individual’s goals and care planning actions.  \\
&& - Support of GP link-working as part of community transformation by identifying service users under a particular GP in joint meetings with Primary Care Community Mental Health Teams (PCMHTs) or direct liaisons. \\ 
&&- Identify social determinants of health outcomes, such as housing stability and employment status, and involve other services and agencies.   \\ \hline

\multirow{3}{3.5cm}{Monitoring and Follow-up}    & \multirow{3}{5cm}{Regularly reviewing patient progress and adjusting treatment plans as needed to ensure the best outcomes.}      &  - Track previous diagnoses and medical history. \\ 
&&- Monitor changes of complexity and risk over time. \\
&&- Medication administration follow-up.                \\
\hline
\multirow{4}{3.5cm}{Documentation and Reporting} & \multirow{4}{5cm}{Maintaining accurate records of patient interactions, treatment plans, and outcomes to track progress and inform future care decisions.}                                                             & - Team caseload summary. \\ 
&&- Review of patient outcomes.\\ 
&&- Improve the documentation of key performance indicators (KPIs). \\
\hline
\end{tabular}
\label{tab:caseload}
\end{table*}

\subsection{Usability and acceptability}
\label{section_engage}

\begin{figure}
	\includegraphics[width=1\columnwidth]{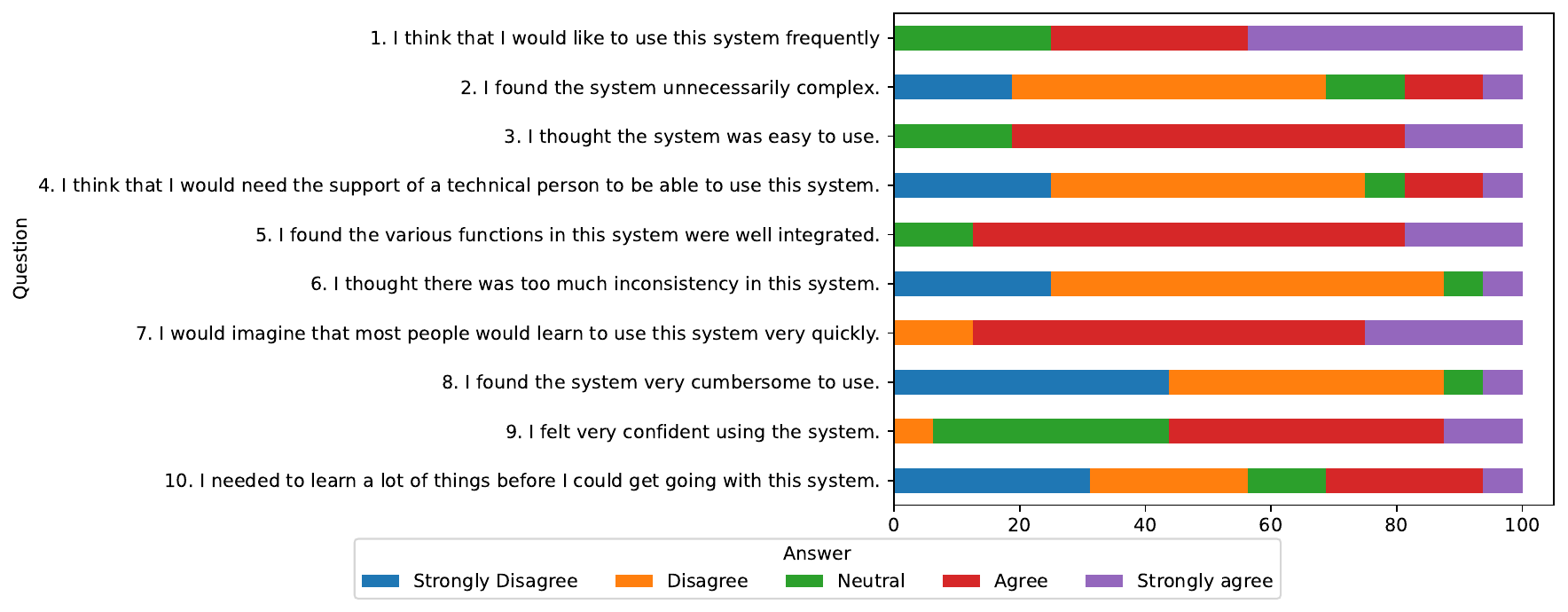}
    \caption{Proportions of various responses to each question in the SUS survey.}
    \label{fig:sus}
\end{figure}

To assess the usability of VIEWER, we used the standard System Usability Scale (SUS), which consists of a 10-item questionnaire with 5 response options ranging from ``Strongly agree'' to ``Strongly disagree'' \cite{brooke1996sus} during its early development stage. \textcolor{black}{16 users completed the questionnaire}, and the overall score was 74/100, indicating ``good'' usability. Figure \ref{fig:sus} presents the detailed responses for each SUS question from the 16 respondents. The positive scores are further supported by the respondents' feedback in \textit{SI} Table S2. Among the responses, 4 users scored the system over 80, indicating potential promoters of the system, while 2 users scored below 55, making them potential detractors. Both potential detractors provided free-text responses stating that they found the system useful and had used it to improve patient care but expressed a need for more training.

\begin{figure*}
	\includegraphics[width=1\columnwidth]{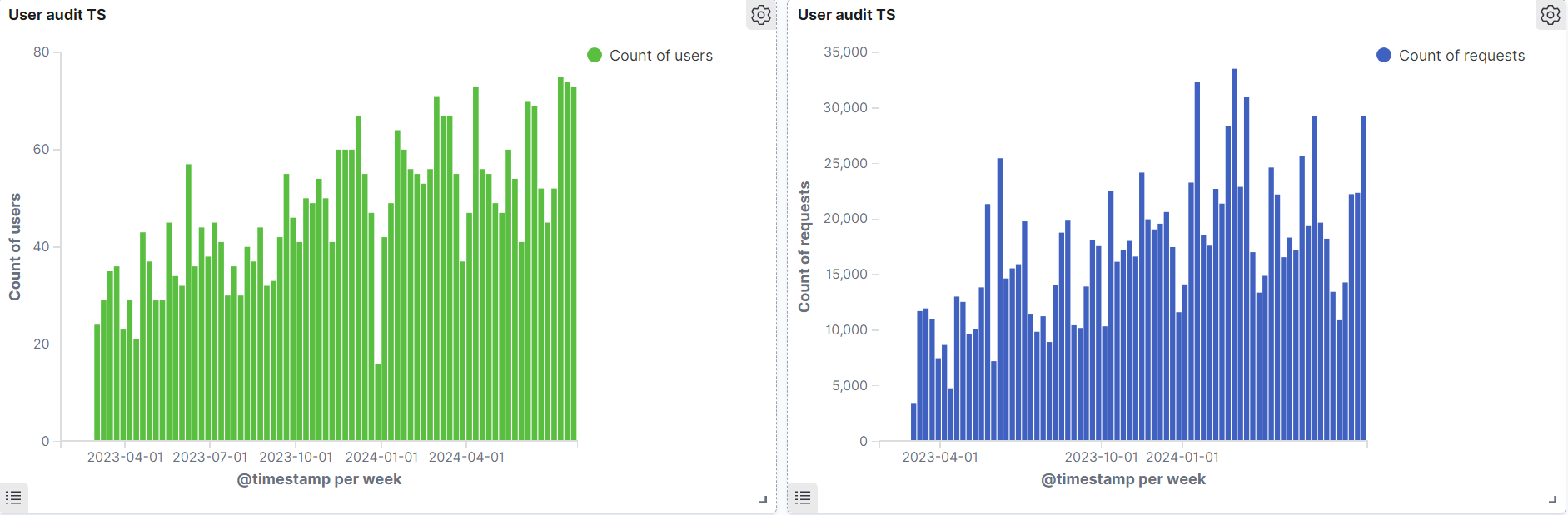}
    \caption{The numbers of users and requests per week in the last 2 years.}
    \label{fig:audit}
\end{figure*}

The acceptability of VIEWER is illustrated by its wide adoption within the Trust. Over 1,000 clinical and managerial staff members, out of approximately 5,000 total staff (including both clinical and non-clinical roles), have used the system over the last two years. VIEWER has handled more than 147 million data requests, with user activity likely to increase, as shown by the trends in Figure \ref{fig:audit}. Due to these positive outcomes, the Trust and the Maudsley Charity have recently committed a 4-year funding to operate the system as a business-as-usual service, aiming to achieve broader and longer-lasting impacts.

\section{Discussion}
This paper presents the design, development, implementation, and evaluation methods of VIEWER: a lightweight, extensible toolkit for rapid creation and deployment of clinical information extraction, analysis and visualisation to enhance data accessibility in various aspects of healthcare delivery.
VIEWER leverages advanced NLP-based information extraction methods to derive clinical insights from diverse EHR data and employs interactive visualisations to effectively communicate these insights, enabling informed decision-making at the point of care. 
Designed to complement, not replace, existing EHR components, VIEWER is built primarily on extensive pre-existing but researcher-focused capability (i.e. CRIS) and open-source technologies such as CogStack, offering a cost-effective enhancement to the functionalities and intelligence of legacy EHR systems. 
This system has demonstrated valuable impacts in various real-world clinical applications within a large mental health Trust.

While many clinical information retrieval and visualisation tools have been developed previously, most have either focused on processing structured tabular data \cite{badgeley2016ehdviz} or specific tasks (e.g, individual patient information summarisation \cite{hirsch2015harvest} or cohort search \cite{callahan2021ace}), and data related to certain diseases \cite{soulakis2015visualizing}. VIEWER advances this field by employing scalable, distributed, NLP-based information extraction pipelines that enable comprehensive search and analysis of both structured and unstructured information for all service users within a Trust. 
While our implementation utilises in-house NLP applications for information extraction, VIEWER supports the integration of open-source or custom NLP models as plugins\footnote{\url{https://github.com/CogStack/CogStack-Pipeline}}. 
Moreover, by leveraging interdisciplinary expertise and participatory design approaches, VIEWER provides a user-friendly interface that seamlessly integrates with existing EHR systems and enables information presented in a useful manner. This capability effectively addresses a broad spectrum of real-world applications, ranging from population health management and service/specialist-specific caseload optimisation to personalised care plans for individual patients. The interactive visualisations also allow for customisation to suit local clinical specialties, workflows and preferences.

VIEWER also offers \textcolor{black}{high potential for interoperability}, allowing for directly deploying on top of existing EHR systems or incorporating into clinical research platforms to facilitate the translation of research into practice. 
In our implementation, we leveraged the clinical research infrastructure in our institution and use VIEWER to facilitate the transformation of research outcomes, particularly in data linkage and NLP pipelines within CRIS, into real-world clinical practices. 
As a result, the extensive set of clinical information extraction and NLP pipelines, developed and refined through multiple CRIS research projects over the last decade, has now been applied to inform clinical decisions directly at the point of care. 
This also enhances the accessibility of clinical research datasets and enables large-scale studies at the population level, e.g. examining outcome disparities across populations \cite{wang2023unraveling}. 
However, when a CRIS-like system is absent, CogStack, the backbone component used for data ingestion and visualisation in VIEWER, provides a set of open-source tools capable of parsing text content from binary documents (e.g. PDFs, scanned images, and Word files) and NLP packages for extracting medical information from clinical text\footnote{\url{https://github.com/CogStack/CogStack-Pipeline}}. 
Indeed, CogStack has been deployed at multiple NHS sites, including both mental and physical healthcare providers in the UK and beyond \cite{jackson-2018-cogstack}.
This offers a high degree of flexibility for easy customisation when integrating VIEWER with various EHR systems directly.

Another notable strength of VIEWER is its extensibility and customizability. Although originally developed within mental healthcare settings, the VIEWER framework is vendor-agnostic and capable of importing data from various relational databases and external systems such as sensor devices and fitness monitors. 
While different settings have varying priorities and workflows, our data model design closely aligns with the Observational Medical Outcomes Partnership (OMOP) Common Data Model, an open community data standard for observational health data sciences and informatics \cite{hripcsak2019book}, with modifications tailored to local settings and project-specific requirements. Also, our interface design adhere to well-established health informatics design methods \cite{clemensen2017participatory,hripcsak2019book} and are general to cover key components in clinical practice. This versatility enables VIEWER to be effectively applied in other settings to enhance data usage and supporting informed clinical decision-making. 
Also, VIEWER supports extensive customisation options, including visualisation coding tools like Vega \cite{satyanarayan2015reactive} and graphical user interfaces (GUIs), as well as integration with external libraries for predictive and statistical learning algorithms available in languages such as Python and R. This customizability empowers both technical and non-technical users to design and refine data visualisations and analytical dashboards. In fact, most visualisations in VIEWER were developed by clinicians rather than technicians, which has enhanced its clinical relevance and facilitated its rapid adoption in practice.

The insights gained and factors contributing to VIEWER's success have led us to make the following recommendations for designing clinical decision support systems (CDSS).

\begin{enumerate}[leftmargin=*]
\item \textbf{Interdisciplinary collaboration}: As highlighted previous guidelines \cite{brandeau2009recommendations}, the success of CDSS hinges on their ability to tackle real-world challenges faced by clinicians. To ensure this, it is essential to co-design a system with input from a broad range of partners with interdisciplinary backgrounds, particularly clinicians and governance committees. 
From our experience in implementing VIEWER, this collaborative approach not only enhances the understanding of practical challenges and helps identify relevant solutions more effectively, but also fosters the involvement of key partners as part of an iterative process to support the creation of practical models that are more likely to be adopted effectively.

\item \textbf{Complexity and usability}: As identified in previous studies, complexity is a major barrier to the adoption of CDSS, which often causes a temporary decline in productivity as users navigate the learning curve of a new system \cite{kruse2018use}. This aligns with our observations when implementating VIEWER. 
To ensure high adoption rates and maximise the system's overall impact, CDSS should strike a good balance between simplicity and complexity, so that it can effectively model real-world scenarios but remain user-friendly and easily interpretable for its intended users \cite{brandeau2009recommendations}. We also find that when achieving simplicity is infeasible, providing adequate training and support is essential to manage the inherent complexity. Moreover, interface components should be organised to fit users' workflows \cite{brown2016interface} and maintain consistent styles to minimise the learning curve. 

\item \textbf{Multi-dimension integration}: Integrating with local EHR systems and aligning with users' workflows are critical for enhancing a CDSS's accessibility and usability. In our implementation of VIEWER, users' most common question was difficulty in logging into the new system. This problem was largely resolved by integrating Single Sign-On (SSO) \cite{de2002single} authentication with the local EHR. 
However, deep integration of a CDSS with existing EHR systems can be challenging due to the required technical infrastructure and staff expertise \cite{melnick2019integrated}.
Therefore, it is important to integrate all components of a CDSS into a cohesive, self-contained system within a sandboxed environment to ensure a smooth user experience during the early stages of implementation.
Also, due to security and safety concerns, real-time interactions (e.g. information entry and modification) are often restricted from being pushed back to the EHR from the CDSS. 
Embedding a hyperlink to allow users to seamlessly navigate between the CDSS and the EHR system can effectively address this issue.

\item \textbf{Iterative feedback}: Regular performance monitoring and evaluation are important to ensure that a CDSS meets clinical needs. 
In the VIEWER system, we developed several visualisations and dashboards to monitor the treads of clinical outcomes over time. 
These monitoring panels not only enable consistent measurements of performance but also feature a user-friendly, interactive interface that allows customisation of parameters to effectively address local needs for performance analyses. 
This creates an electronic audit and feedback cycle which starts from informing goal setting, streamlining interventions to automatically collecting and analyzing data, leading to iterative feedback to optimise performance. 
These feedback cycles are also useful for motivating behavior changes and system adoption, particularly when the feedback directly supports clinical behaviors \cite{brown2019clinical}.

\item \textbf{Transparent communication}: A new system may initially fall short of users' expectations and require their engagement for further improvement. 
While promptly responding to users' requests and feedback is vital to maintain their engagement, active listening and transparent communication are essential to maximise the outcomes of such engagement. 
In our interactions with VIEWER users, we highlighted not only the strengths of our solutions but also their limitations, encouraging users to contribute to resolving these issues and refining the system. 
This approach fosters a sense of ``ownership'' over the feedback process, making users feel more invested in the improvements. Such ownership-driven feedback leverages users' autonomy and internal motivation to enhance the system, rather than feeling subjected to external policies or directives \cite{brown2019clinical}. 
\end{enumerate}

This includes incorporating robust privacy-preserving techniques, such as differential privacy, and ensuring that all stakeholders (clinicians, patients, and developers) are properly informed about how their data is being used and processed. Ethical considerations, especially regarding mental health data, will be crucial as we continue to develop and scale such technologies.

VIEWER and this study of its use also have limitations and requires further improvement. 
First, although a range of NLP applications have been used to extract useful information from clinical text, which contains the majority of patient information, this task remains challenging. The primary difficulties stem from the extensive use of medical terminology, grammatical inaccuracies, and the sensitivity of clinical data which results in limited availability of data for training large language models \cite{wang2018clinical}. Our users have reported that temporality (distinguishing past, present, or future events), certainty (differentiating confirmed and hypothetical events), and negation are the most common sources of errors in NLP extraction results, highlighting areas that require further improvement \cite{mahajan2021toward}. As a contingency solution, VIEWER provides text snippets that offer complete context for each extracted event, allowing users to interpret and validate the NLP results effectively. Moreover, real-time extraction of information and aggregation of NLP-extracted information to construct care trajectories proves useful but also require further development. \textcolor{black}{Given the sensitive nature of mental healthcare data, it is essential to prioritize patient confidentiality and privacy in the development of these technologies. This requires strict adherence to ethical guidelines, such as data anonymization, and compliance with regulations like the GDPR and the NHS Data Security and Protection Toolkit.}

Second, beyond functionality, the sustainability of healthcare informatics solutions—encompassing aspects such as maintainability, scalability, and interoperability—is crucial for enhancing the quality and affordability of healthcare delivery. To fully understand the benefits and limitations of VIEWER, it is important to compare it with other open-source or commercial solutions that offer similar capabilities. A more extensive comparison, based on longer and broader usage, would provide valuable insights into whether the findings are widely applicable.
Third, VIEWER has primarily been used by clinicians at SLaM. However, it would be beneficial to make VIEWER's patient chart features available to individual patients, enabling them to monitor their own health metrics and input relevant health data. To explore this, we would need authorisation to grant patient access to the system, as well as technical support and data integration services to connect with other patient portals.

Finally, VIEWER has only been deployed at SLaM so far and thus our evaluation focused on findings in a single institution. However, as VIEWER was designed to be EHR-agnostic, porting it to another EHR system is feasible. We are eager to deploy VIEWER in other institutions and assess the generalizability of our findings. 
Moreover, \textcolor{black}{this paper has primarily focused on outlining the technical components and development process of VIEWER, and our current evaluation relies on observational research methods. To formally assess the effectiveness of our informatics intervention and determine the factors contributing to its success, such as the types of use cases and participant characteristics, including backgrounds, demographics, and areas of expertise, across various teams and settings, randomised controlled trials are needed}. 

\section{Conclusion}
As EHR data rapidly increases, there is an urgent need to enhance data accessibility and interpretability across various aspects of healthcare delivery, such as population health management, caseload optimisation, patient monitoring and personalisation of care. Through interdisciplinary collaboration, we developed VIEWER: a cost-effective, open-source, and extensible toolkit designed for the rapid design, development, and implementation of clinical information retrieval, analysis, and visualisation to meet this need. 
Deploying VIEWER in one of the largest mental healthcare provider in the UK further enables us to explore the use of patient data visual-analytics as an informatics intervention to enhance patient care in real-world clinical settings. Our implementation offers valuable insights into collaborating with clinicians to iteratively refine and optimise the interface and content of a health informatics solution for better patient care.


\section*{Funding Statement}
This study was funded by: (a) Maudsley Charity, (b) the National Institute for Health Research (NIHR) Biomedical Research Centre at South London and Maudsley NHS Foundation Trust and King’s College London; (c) Health Data Research UK, which is funded by the UK Medical Research Council, Engineering and Physical Sciences Research Council, Economic and Social Research Council, Department of Health and Social Care (England), Chief Scientist Office of the Scottish Government Health and Social Care Directorates, Health and Social Care Research and Development Division (Welsh Government), Public Health Agency (Northern Ireland), British Heart Foundation and Wellcome Trust; (d) The BigData@Heart Consortium, funded by the Innovative Medicines Initiative-2 Joint Undertaking under grant agreement No. 116074. This Joint Undertaking receives support from the European Union’s Horizon 2020 research and innovation programme and EFPIA; it is chaired, by DE Grobbee and SD Anker, partnering with 20 academic and industry partners and ESC; and (e) The National Institute for Health Research University College London Hospitals Biomedical Research Centre. These funding bodies had no role in the design of the study, collection and analyses. The views expressed are those of the author(s) and not necessarily those of the NHS, the NIHR or the Department of Health. 
RS is part-funded by: i) the National Institute for Health Research (NIHR) Biomedical Research Centre at the South London and Maudsley NHS Foundation Trust and King’s College London; ii) an NIHR Senior Investigator Award; iii) the National Institute for Health Research (NIHR) Applied Research Collaboration South London (NIHR ARC South London) at King’s College Hospital NHS Foundation Trust; iv) the DATAMIND HDR UK Mental Health Data Hub (MRC grant MR/W014386). TW was also funded by Early Career Research Award from the Institute of Psychiatry, Psychology \& Neuroscience.

\section*{Competing Interests Statement}
RS declares research support received in the last 36 months from Janssen, GSK and Takeda, and royalties from Oxford University Press.

\section*{Contributorship Statement}
TW, DC, YJM, MB, and RH contributed to the conceptualization, methodology design, data acquisition, analysis, interpretation, and software development. Data extraction and preprocessing were carried out by MB, TW, YJM, and DK, with guidance from DC and RH. Method development was led by TW, DC, and YJM under the guidance of CP, CDP, TS, BA, SM, ZK, MD, AR, RS, PM, RD, and RH. Funding acquisition and supervision were supported by AR, RS, PM, and RD. TW drafted the initial manuscript and implemented revisions, while all authors participated in reviewing and editing the final version.

\section*{Acknowledgements}
The authors would like to express their gratitude to Damian Larkin, Ian Cutting-Jones, Omar Rayner-Andrews, Percy Chinouya, Juliet Hurn, Abigail Bennett, Justine Chow, Man Yuen, Shubhra Mace, Sandra Batchelor, Isaac Enakimio, and many others for their valuable advice and contributions to the project. 

\section*{Data Availability}
All data utilized in this study are derived from patient records within SLaM and are not publicly accessible. Details about the software and tools employed in the research are provided within the article, including access links where available or upon request.

\bibliographystyle{vancouver}
\bibliography{template} 

\begin{thebibliography}{10}

\bibitem{mcdonald1976protocol}
McDonald CJ.
\newblock Protocol-based computer reminders, the quality of care and the non-perfectability of man.
\newblock New England Journal of Medicine. 1976;295(24):1351-5.

\bibitem{singh2013information}
Singh H, Spitzmueller C, Petersen NJ, Sawhney MK, Sittig DF.
\newblock Information overload and missed test results in electronic health record--based settings.
\newblock JAMA internal medicine. 2013;173(8):702-4.

\bibitem{nijor2022patient}
Nijor S, Rallis G, Lad N, Gokcen E.
\newblock Patient safety issues from information overload in electronic medical records.
\newblock Journal of Patient Safety. 2022;18(6):e999-e1003.

\bibitem{adler2011survey}
Adler-Milstein J, Bates DW, Jha AK.
\newblock A survey of health information exchange organizations in the United States: implications for meaningful use.
\newblock Annals of internal medicine. 2011;154(10):666-71.

\bibitem{skou2022multimorbidity}
Skou ST, Mair FS, Fortin M, Guthrie B, Nunes BP, Miranda JJ, et~al.
\newblock Multimorbidity.
\newblock Nature Reviews Disease Primers. 2022;8(1):48.

\bibitem{silverman2015american}
Silverman JJ, Galanter M, Jackson-Triche M, Jacobs DG, Lomax JW, Riba MB, et~al.
\newblock The American Psychiatric Association practice guidelines for the psychiatric evaluation of adults.
\newblock American Journal of Psychiatry. 2015;172(8):798-802.

\bibitem{hirsch2015harvest}
Hirsch JS, Tanenbaum JS, Lipsky~Gorman S, Liu C, Schmitz E, Hashorva D, et~al.
\newblock HARVEST, a longitudinal patient record summarizer.
\newblock Journal of the American Medical Informatics Association. 2015;22(2):263-74.

\bibitem{badgeley2016ehdviz}
Badgeley MA, Shameer K, Glicksberg BS, Tomlinson MS, Levin MA, McCormick PJ, et~al.
\newblock EHDViz: clinical dashboard development using open-source technologies.
\newblock BMJ open. 2016;6(3):e010579.

\bibitem{west2015innovative}
West VL, Borland D, Hammond WE.
\newblock Innovative information visualization of electronic health record data: a systematic review.
\newblock Journal of the American Medical Informatics Association. 2015;22(2):330-9.

\bibitem{wang2022ehr}
Wang Q, Laramee RS.
\newblock EHR STAR: the state-of-the-art in interactive EHR visualization.
\newblock In: Computer graphics forum. vol.~41. Wiley Online Library; 2022. p. 69-105.

\bibitem{powsner1994graphical}
Powsner SM, Tufte ER.
\newblock Graphical summary of patient status.
\newblock Lancet. 1994;344(8919):386-9.

\bibitem{polhemus2022data}
Polhemus A, Novak J, Majid S, Simblett S, Morris D, Bruce S, et~al.
\newblock Data visualization for chronic neurological and mental health condition self-management: systematic review of user perspectives.
\newblock JMIR mental health. 2022;9(4):e25249.

\bibitem{shahar1999intelligent}
Shahar Y, Cheng C.
\newblock Intelligent visualization and exploration of time-oriented clinical data.
\newblock In: Proceedings of the 32nd Annual Hawaii International Conference on Systems Sciences. 1999. HICSS-32. Abstracts and CD-ROM of Full Papers. IEEE; 1999. p. 12-pp.

\bibitem{plaisant2003lifelines}
Plaisant C, Mushlin R, Snyder A, Li J, Heller D, Shneiderman B.
\newblock LifeLines: using visualization to enhance navigation and analysis of patient records.
\newblock In: The craft of information visualization. Elsevier; 2003. p. 308-12.

\bibitem{feblowitz2011summarization}
Feblowitz JC, Wright A, Singh H, Samal L, Sittig DF.
\newblock Summarization of clinical information: a conceptual model.
\newblock Journal of biomedical informatics. 2011;44(4):688-99.

\bibitem{callahan2021ace}
Callahan A, Polony V, Posada JD, Banda JM, Gombar S, Shah NH.
\newblock ACE: the Advanced Cohort Engine for searching longitudinal patient records.
\newblock Journal of the American Medical Informatics Association. 2021;28(7):1468-79.

\bibitem{elshehaly2020qualdash}
Elshehaly M, Randell R, Brehmer M, McVey L, Alvarado N, Gale CP, et~al.
\newblock QualDash: Adaptable generation of visualisation dashboards for healthcare quality improvement.
\newblock IEEE Transactions on Visualization and Computer Graphics. 2020;27(2):689-99.

\bibitem{soulakis2015visualizing}
Soulakis ND, Carson MB, Lee YJ, Schneider DH, Skeehan CT, Scholtens DM.
\newblock Visualizing collaborative electronic health record usage for hospitalized patients with heart failure.
\newblock Journal of the American Medical Informatics Association. 2015;22(2):299-311.

\bibitem{chishtie2022interactive}
Chishtie J, Bielska IA, Barrera A, Marchand JS, Imran M, Tirmizi SFA, et~al.
\newblock Interactive visualization applications in population health and health services research: systematic scoping review.
\newblock Journal of medical Internet research. 2022;24(2):e27534.

\bibitem{huang2015novel}
Huang CW, Syed-Abdul S, Jian WS, Iqbal U, Nguyen PA, Lee P, et~al.
\newblock A novel tool for visualizing chronic kidney disease associated polymorbidity: a 13-year cohort study in Taiwan.
\newblock Journal of the American Medical Informatics Association. 2015;22(2):290-8.

\bibitem{franklin2017dashboard}
Franklin A, Gantela S, Shifarraw S, Johnson TR, Robinson DJ, King BR, et~al.
\newblock Dashboard visualizations: Supporting real-time throughput decision-making.
\newblock Journal of biomedical informatics. 2017;71:211-21.

\bibitem{caban2015visual}
Caban JJ, Gotz D.
\newblock Visual analytics in healthcare--opportunities and research challenges.
\newblock Journal of the American Medical Informatics Association. 2015;22(2):260-2.

\bibitem{chung2020role}
Chung Y, Bagheri N, Salinas-Perez JA, Smurthwaite K, Walsh E, Furst M, et~al.
\newblock Role of visual analytics in supporting mental healthcare systems research and policy: A systematic scoping review.
\newblock International Journal of Information Management. 2020;50:17-27.

\bibitem{salvador2006framework}
Salvador-Carulla L, Haro J, Ayuso-Mateos J.
\newblock A framework for evidence-based mental health care and policy.
\newblock Acta Psychiatrica Scandinavica. 2006;114:5-11.

\bibitem{roe2022patient}
Roe D, Mazor Y, Gelkopf M.
\newblock Patient-reported outcome measurements (PROMs) and provider assessment in mental health: a systematic review of the context of implementation.
\newblock International Journal for Quality in Health Care. 2022;34(Supplement\_1):ii28-39.

\bibitem{kadra2015extracting}
Kadra G, Stewart R, Shetty H, Jackson RG, Greenwood MA, Roberts A, et~al.
\newblock Extracting antipsychotic polypharmacy data from electronic health records: developing and evaluating a novel process.
\newblock BMC psychiatry. 2015;15(1):1-7.

\bibitem{ibrahim2020rapid}
Ibrahim H, Sorrell S, Nair SC, Al~Romaithi A, Al~Mazrouei S, Kamour A.
\newblock Rapid development and utilization of a clinical intelligence dashboard for frontline clinicians to optimize critical resources during COVID-19.
\newblock Acta Informatica Medica. 2020;28(3):209.

\bibitem{schall2017usability}
Schall~Jr MC, Cullen L, Pennathur P, Chen H, Burrell K, Matthews G.
\newblock Usability evaluation and implementation of a health information technology dashboard of evidence-based quality indicators.
\newblock CIN: Computers, Informatics, Nursing. 2017;35(6):281-8.

\bibitem{lim2022toward}
Lim HC, Austin JA, Van Der~Vegt AH, Rahimi AK, Canfell OJ, Mifsud J, et~al.
\newblock Toward a learning health care system: a systematic review and evidence-based conceptual framework for implementation of clinical analytics in a digital hospital.
\newblock Applied clinical informatics. 2022;13(02):339-54.

\bibitem{stewart2009south}
Stewart R, Soremekun M, Perera G, Broadbent M, Callard F, Denis M, et~al.
\newblock The South London and Maudsley NHS foundation trust biomedical research centre (SLAM BRC) case register: development and descriptive data.
\newblock BMC psychiatry. 2009;9(1):1-12.

\bibitem{jackson-2018-cogstack}
Jackson R, Kartoglu I, Stringer C, Gorrell G, Roberts A, Song X, et~al.
\newblock CogStack-experiences of deploying integrated information retrieval and extraction services in a large National Health Service Foundation Trust hospital.
\newblock BMC medical informatics and decision making. 2018;18(1):1-13.

\bibitem{jensen2018bridging}
Jensen CM, Overgaard S, Wiil UK, Smith AC, Clemensen J.
\newblock Bridging the gap: a user-driven study on new ways to support self-care and empowerment for patients with hip fracture.
\newblock SAGE open medicine. 2018;6:2050312118799121.

\bibitem{sedlmair2012design}
Sedlmair M, Meyer M, Munzner T.
\newblock Design study methodology: Reflections from the trenches and the stacks.
\newblock IEEE transactions on visualization and computer graphics. 2012;18(12):2431-40.

\bibitem{clemensen2017participatory}
Clemensen J, Rothmann MJ, Smith AC, Caffery LJ, Danbjorg DB.
\newblock Participatory design methods in telemedicine research.
\newblock Journal of telemedicine and telecare. 2017;23(9):780-5.

\bibitem{wang2023unraveling}
Wang T, Codling D, Bhugra D, Msosa Y, Broadbent M, Patel R, et~al.
\newblock Unraveling ethnic disparities in antipsychotic prescribing among patients with psychosis: a retrospective cohort study based on electronic clinical records.
\newblock Schizophrenia Research. 2023;260:168-79.

\bibitem{noauthor_cris_nodate}
{CRIS} {Natural} {Language} {Processing};.
\newblock Available from: \url{https://www.maudsleybrc.nihr.ac.uk/facilities/clinical-record-interactive-search-cris/cris-natural-language-processing/}.

\bibitem{noauthor_cris_nodate-1}
{CRIS} {Data} {Linkages};.
\newblock Available from: \url{https://maudsleybrc.nihr.ac.uk/facilities/clinical-record-interactive-search-cris/cris-data-linkages/}.

\bibitem{cunningham2013getting}
Cunningham H, Tablan V, Roberts A, Bontcheva K.
\newblock Getting more out of biomedical documents with GATE's full lifecycle open source text analytics.
\newblock PLoS Comput Biol. 2013;9(2):e1002854.

\bibitem{perera2016cohort}
Perera G, Broadbent M, Callard F, Chang CK, Downs J, Dutta R, et~al.
\newblock Cohort profile of the South London and Maudsley NHS Foundation trust biomedical research centre (SLAM BRC) case register: current status and recent enhancement of an electronic mental health Record-derived data resource.
\newblock BMJ open. 2016;6(3).

\bibitem{sso}
SSO and LDAP Authentication;.
\newblock Available from: \url{https://archive.ph/20140523114521/http://www.authenticationworld.com/Single-Sign-On-Authentication/SSOandLDAP.html}.

\bibitem{thornicroft2011physical}
Thornicroft G.
\newblock Physical health disparities and mental illness: the scandal of premature mortality.
\newblock The British Journal of Psychiatry. 2011;199(6):441-2.

\bibitem{john2018premature}
John A, McGregor J, Jones I, Lee SC, Walters JT, Owen MJ, et~al.
\newblock Premature mortality among people with severe mental illness—New evidence from linked primary care data.
\newblock Schizophrenia research. 2018;199:154-62.

\bibitem{chesney2014risks}
Chesney E, Goodwin GM, Fazel S.
\newblock Risks of all-cause and suicide mortality in mental disorders: a meta-review.
\newblock World psychiatry. 2014;13(2):153-60.

\bibitem{smith2013schizophrenia}
Smith DJ, Langan J, McLean G, Guthrie B, Mercer SW.
\newblock Schizophrenia is associated with excess multiple physical-health comorbidities but low levels of recorded cardiovascular disease in primary care: cross-sectional study.
\newblock BMJ open. 2013;3(4).

\bibitem{wang2022patient}
Wang T, Bendayan R, Msosa Y, Pritchard M, Roberts A, Stewart R, et~al.
\newblock Patient-centric characterization of multimorbidity trajectories in patients with severe mental illnesses: A temporal bipartite network modeling approach.
\newblock Journal of Biomedical Informatics. 2022;127:104010.

\bibitem{moncrieff2016results}
Moncrieff J, Azam K, Johnson S, Marston L, Morant N, Darton K, et~al.
\newblock Results of a pilot cluster randomised trial of the use of a Medication Review Tool for people taking antipsychotic medication.
\newblock BMC psychiatry. 2016;16:1-11.

\bibitem{kane2013non}
Kane JM, Kishimoto T, Correll CU.
\newblock Non-adherence to medication in patients with psychotic disorders: epidemiology, contributing factors and management strategies.
\newblock World psychiatry. 2013;12(3):216-26.

\bibitem{brooke1996sus}
Brooke J, et~al.
\newblock SUS-A quick and dirty usability scale.
\newblock Usability evaluation in industry. 1996;189(194):4-7.

\bibitem{hripcsak2019book}
Hripcsak G, Ryan P, Madigan D, Kostka K, Schuemie M, DeFalco F. The Book of OHDSI: Observational Health Data Sciences and Informatics. OHDSI; 2019.

\bibitem{satyanarayan2015reactive}
Satyanarayan A, Russell R, Hoffswell J, Heer J.
\newblock Reactive vega: A streaming dataflow architecture for declarative interactive visualization.
\newblock IEEE transactions on visualization and computer graphics. 2015;22(1):659-68.

\bibitem{brandeau2009recommendations}
Brandeau ML, McCoy JH, Hupert N, Holty JE, Bravata DM.
\newblock Recommendations for modeling disaster responses in public health and medicine: a position paper of the society for medical decision making.
\newblock Medical Decision Making. 2009;29(4):438-60.

\bibitem{kruse2018use}
Kruse CS, Stein A, Thomas H, Kaur H.
\newblock The use of electronic health records to support population health: a systematic review of the literature.
\newblock Journal of medical systems. 2018;42(11):214.

\bibitem{brown2016interface}
Brown B, Balatsoukas P, Williams R, Sperrin M, Buchan I.
\newblock Interface design recommendations for computerised clinical audit and feedback: hybrid usability evidence from a research-led system.
\newblock International journal of medical informatics. 2016;94:191-206.

\bibitem{de2002single}
De~Clercq J.
\newblock Single sign-on architectures.
\newblock In: International Conference on Infrastructure Security. Springer; 2002. p. 40-58.

\bibitem{melnick2019integrated}
Melnick ER, Holland WC, Ahmed OM, Ma AK, Michael SS, Goldberg HS, et~al.
\newblock An integrated web application for decision support and automation of EHR workflow: a case study of current challenges to standards-based messaging and scalability from the EMBED trial.
\newblock JAMIA open. 2019;2(4):434-9.

\bibitem{brown2019clinical}
Brown B, Gude WT, Blakeman T, van~der Veer SN, Ivers N, Francis JJ, et~al.
\newblock Clinical Performance Feedback Intervention Theory (CP-FIT): a new theory for designing, implementing, and evaluating feedback in health care based on a systematic review and meta-synthesis of qualitative research.
\newblock Implementation Science. 2019;14:1-25.

\bibitem{wang2018clinical}
Wang Y, Wang L, Rastegar-Mojarad M, Moon S, Shen F, Afzal N, et~al.
\newblock Clinical information extraction applications: a literature review.
\newblock Journal of biomedical informatics. 2018;77:34-49.

\bibitem{mahajan2021toward}
Mahajan D, Liang JJ, Tsou CH.
\newblock Toward understanding clinical context of medication change events in clinical narratives.
\newblock In: AMIA Annual Symposium Proceedings. vol. 2021. American Medical Informatics Association; 2021. p. 833.

\end{thebibliography}






\appendix
\renewcommand{\thefigure}{S\arabic{figure}} 
\renewcommand{\thetable}{S\arabic{table}}   
\setcounter{figure}{0} 
\setcounter{table}{0}  

\section{Appendix: Supplementary Information}

\begin{figure*}
	\includegraphics[width=1\columnwidth]{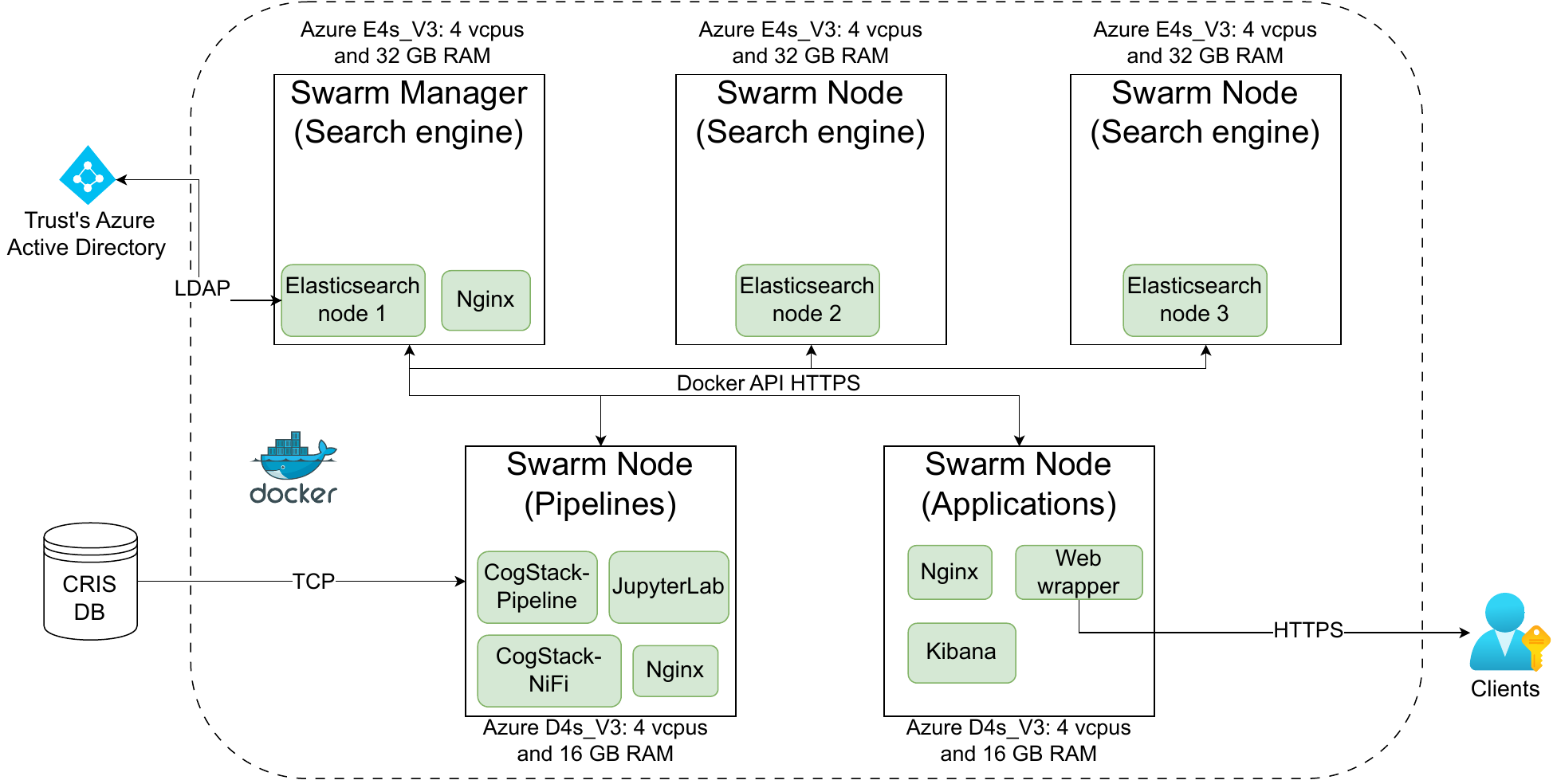}
    \caption[Docker-based deployment of CogStack]{A distributed deployment of CogStack based on Docker Swarm. Five Azure virtual machines host three types of containerised micro-services: 1) data storage, index and search engine, 2) data ingestion and analytical pipelines, and 3) interface applications.}
    \label{fig:docker}
\end{figure*}

\begin{table*}
\scriptsize
\centering
\caption[Descriptive statistics of data models]{Descriptive statistics of data models.}
\begin{tabular}{lp{6cm}lr}
\hline
Table/Index                                & Description                                 & Column Count & Row Count  \\
\hline
BMI catalogue              & BMI values of each patient over time, including those recorded in both structured fields and text notes extracted using NLP methods.                                        & 11           & 399,277      \\
Psychology episodes       & Psychological therapy episodes for each patient over time.              & 39           & 84,224       \\
CORE\_OM\ scores                   & CORE Outcome Measure (CORE-OM) scores for each patient over time.        & 70           & 103,528      \\
Medication catalogue & Each patient's medication records over time, including structured fields and records in text notes extracted using NLP.                               & 29           & 18,929,237    \\
Test results    & Lab results of patients over time.                                 & 14           & 704,327      \\
Contact catalogue                   & A patient's service contacts (e.g. bed days, community care and emergency services) over time.                           & 17           & 17,373,156    \\
Clinician caseload        & All patients that a staff member is currently caring for.                          & 20           & 46,892       \\
Diagnosis catalogue                  & Diagnosis records of each patient over time.                                  & 18           & 1,061,482     \\
Full caseload        & Caseload episodes for each patient, including latest information about patient details, referrals, care teams, diagnoses, treatments, and outcome observations such as lab tests and DIALOG.                                 & 625          & 1,611,007     \\
Full caseload snapshots                          & All snapshots of the caseload table for monitoring data over time. & 625          & 510,992,515 \\
\hline
\end{tabular}

\label{tab.data}
\end{table*}

\begin{table}
\scriptsize
\caption[User feedback]{Respondent feedback on the overall use of VIEWER. }
\begin{tabular}{p{14cm}}
\hline
\hline
``It has helped with improving patient's physical health care as it is able to identify what physical health checks a client has not had therefore we are able to identify these easier and put in place interventions quicker.''                                                                                                                               \\
\\
``It has definitely stimulated interest to use in practice. It is a helpful tool in presenting, reviewing and auditing.''                                                                                                                                                                                                                                      \\\\
``It definitely helpful in community transformation project, in identifying the most vulnerable patients and demographic. The GP practice which to focus attention as we move into neighbourhood workings, the number of patients on depot medication where to support in the transition to primary care.''                                                     \\\\
``Helps to identify physical health needs and help determine what interventions need to be offered and signposted to external agencies. Helps identify what anti-psychotics have been offered and whether clozapine should be considered as an option. It helps collating information on ePJS (the Trust's EHR system) which saves time - CC's don't have to manually scrutinise notes.'' \\\\
``I was using this to visualise ethnicity data to support a policy lead in conducting an Equality Impact Assessment on the Clozapine policy. This will hopefully inform further ethnicity analysis, discussion and action in relation to this policy and its implementation.''       \\                    \hline                                                                
\end{tabular}
\label{tab:feed}
\end{table}

\begin{figure}
	\includegraphics[width=1\columnwidth]{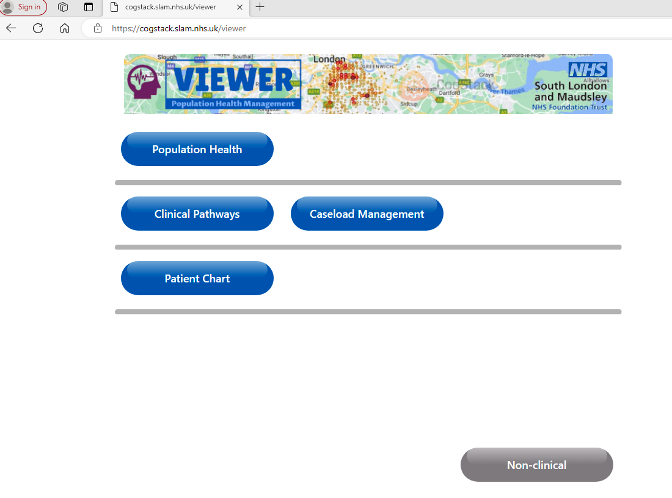}
    \caption[VIEWER home page]{VIEWER home page. Dashboards are organised into four main categories, namely ``Population Health'', ``Clinical Pathways'', ``Caseload Management'' and ``Patient Chart''. Each category employs a hierarchical structure to facilitate clear navigation among sub-sections, such as different diagnoses under ``Clinical Pathways''. The ``Non Clinical''/``Clinical'' button in the right bottom allows users to switch between de-identified and identifiable versions.}
    \label{fig:panel}
\end{figure}

\end{document}